\documentclass[11pt]{article}
\usepackage{moriond,epsfig}

\def\be{\begin{equation}}
\def\ee{\end{equation}}
\def\bea{\begin{eqnarray}}
\def\eea{\end{eqnarray}}

\begin{document}
\vspace*{4cm}
\title{SINGLE TOP PRODUCTION IN THE $Wt$ MODE WITH MC@NLO}

\author{Chris D. White}

\address{Nikhef, Kruislaan 409, 1098SJ Amsterdam, The Netherlands}

\maketitle\abstracts{We consider whether $Wt$ production can be considered 
as a distinct production process at the LHC, separate from top pair production with which 
it interferes. We argue that this problem can be meaningfully addressed in an MC@NLO calculation, 
and give two definitions of the $Wt$ mode whose difference measures the degree of interference
between $Wt$ and top pair production. These are then implemented in the MC@NLO software framework, and
results given which demonstrate that it is indeed legitimate to isolate the $Wt$ process, 
subject to adequate cuts.
}

\section{Introduction}
Top physics is an important research area, and is of theoretical interest given the proximity
of the top quark mass to the energy scale associated with electroweak symmetry breaking. Given
one expects new physics to explain the nature of this symmetry breaking, it follows that the top 
quark sector can be a sensitive probe of new physics effects. Furthermore, the forthcoming
Large Hadron Collider will offer unprecedented rates for top quark production, making detailed
theoretical understanding of its properties essential.

\begin{figure}[h]
\begin{center}
\psfig{figure=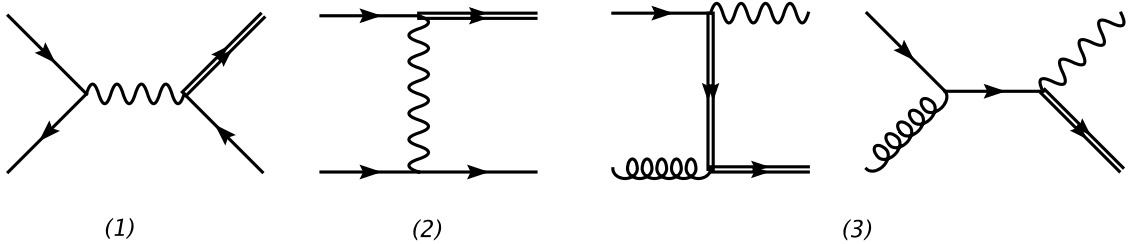,height=1.35in}
\caption{The three SM single top production modes at LO: (1) s-channel; (2) t-channel; (3) Wt production.
\label{LO}}
\end{center}
\end{figure}

Of particular interest is single top production, which efficiently isolates the electroweak 
sector of the Standard Model (SM). The total leading-order (LO) cross-section for single top production
is approximately 320pb, which is somewhat less than the top pair cross-section ($\simeq 830$pb), but still
significant. There are three production modes for single top quarks in the SM, named the $s-$, $t-$ and $Wt$ channels, and shown at LO in figure~\ref{LO}.
The first two are well-understood theoretically, and the latter less so. Despite this, the $Wt$ channel makes 
up around 20\% of the total single top cross-section.

\begin{figure}[h]
\begin{center}
\psfig{figure=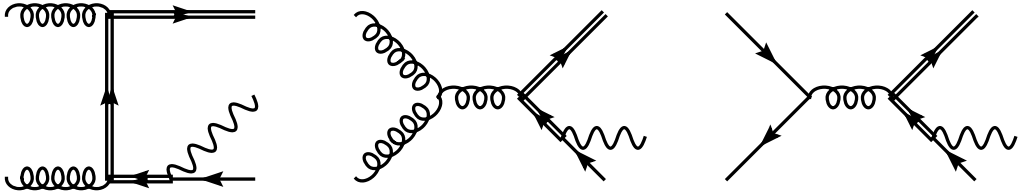,height=1.0in}
\caption{Subset of diagrams contributing to $Wt$ production at NLO in QCD.
\label{doubleres}}
\end{center}
\end{figure}

The $Wt$ mode is well-defined at LO, but at NLO in QCD receives large corrections 
due to diagrams such as those shown in figure~\ref{doubleres}.
These can be interpreted as top pair production, followed by decay of the antitop. The large contributions
 occur when the invariant mass of the final state $W$ and $b$ approaches the top quark mass, 
and the intermediate top quark propagator becomes resonant. The question then arises of whether it is still
possible to consider $Wt$ as a well-defined production process, or whether interference with top pair production
renders this approach untenable.

One consistent procedure is to only consider given final states (namely $WWb$ and $WWb\bar{b}$, see 
e.g.~\cite{Kersevan:2006fq}). One then includes all Feynman diagrams that give rise to each final state, 
truncating the perturbation expansion at ${\cal O}(\alpha_S^2\alpha_{EW})$. 
However, in that case NLO QCD corrections to top pair production are absent,
compromising the theoretical accuracy of this approach. Alternatively, one may choose to regard $Wt$ as a 
well-defined production mode, {\it subject to adequate cuts}. The aim is to give a definition of $Wt$ production
such that one may consistently combine event samples of $Wt$ and $t\bar{t}$ events in any given analysis, without
worrying about interference between them. Provided this can be shown to be consistent
in a wide enough area of phase space, this is a useful approximation to the underlying physics. Firstly,
one has a means of efficiently generating $Wt$-like events for use in Monte Carlo analyses. 
Secondly, full NLO QCD corrections can be included in both the top pair and $Wt$ production modes. 

Previous definitions of the $Wt$ mode were defined (up to NLO level) in~\cite{Belyaev:1998dn,Tait:1999cf,Campbell:2005bb}. 
However, for a definition of the $Wt$ mode to work in an experimental setting, it must be applicable 
within a parton shower context. Because the interference problem only occurs at NLO in the $Wt$ process, this
necessitates the use of an MC@NLO, and we adopt the framework of~\cite{Frixione:2002ik}. 
\section{The $Wt$ Mode in MC@NLO}
We give two definitions of the $Wt$ mode~\cite{Frixione:2008yi}, such that the difference between them measures 
the interference between $Wt$ and $t\bar{t}$ production. These are called diagram removal (DR) 
and diagram subtraction (DS), and can be summarised as follows.

In DR, one removes all diagrams containing an intermediate $t\bar{t}$ pair at the amplitude level, completely 
eradicating the interference with $t\bar{t}$. This has the disadvantage that it breaks QCD gauge invariance,
although this does not seem to be a problem in practice (see~\cite{Frixione:2008yi} for a detailed discussion).

\begin{figure}[h]
\begin{center}
\psfig{figure=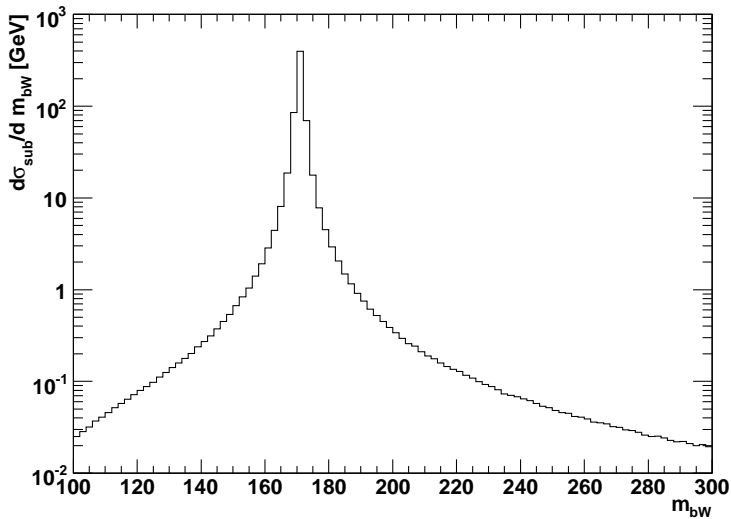,height=3.0in}
\caption{The local subtraction term used in the DS definition of the $Wt$ mode.
\label{sub}}
\end{center}
\end{figure}

In DS, one modifies the differential cross-section with a local subtraction term which removes the resonant 
top pair contribution. This is gauge invariant, and because the definition is at the squared amplitude level 
the interference term with $Wt$ production is still present. The local subtraction term is defined as follows:
\begin{equation}
d\sigma_{sub}=|\tilde{A}(tW\bar{b})_{t\bar{t}}|^2\times\frac{f_{BW}(m_{bW})}{f_{BW}(m_t)},
\label{dsigsub}
\end{equation}
where $\tilde{A}(tW\bar{b})_{t\bar{t}}$ is the amplitude for $tW\bar{b}$ production coming from $t\bar{t}$-like 
diagrams, and the kinematics are reshuffled to place the $\bar{t}$ on-shell. This is then damped by a ratio of 
Breit-Wigner functions $f_{BW}$ when the invariant mass $m_{bW}$ lies
away from the top mass $m_t$ (for more details, see~\cite{Frixione:2008yi}). The subtraction term has the desired 
properties of being strongly peaked around $m_{bW}\simeq m_t$ (with the amplitude at that value 
corresponding to the size of the resonant top pair contribution), and steeply falling as $m_{bW}$ moves away from $m_t$.
A plot of the subtraction term, as a function of $m_{bW}$, is shown in figure~\ref{sub}, and it indeed has these properties.

Both the DR and DS definitions have been implemented in the MC@NLO event generator, and can be found in the latest release~\cite{Frixione:2008ym}. Furthermore, spin correlations have been implemented using the method of~\cite{Frixione:2007zp}. Along with the $s-$ and $t-$ channel modes~\cite{Frixione:2005vw}, this then completes the description of the three single top production modes in MC@NLO.
\section{Results}
\begin{figure}[h]
\begin{center}
\psfig{figure=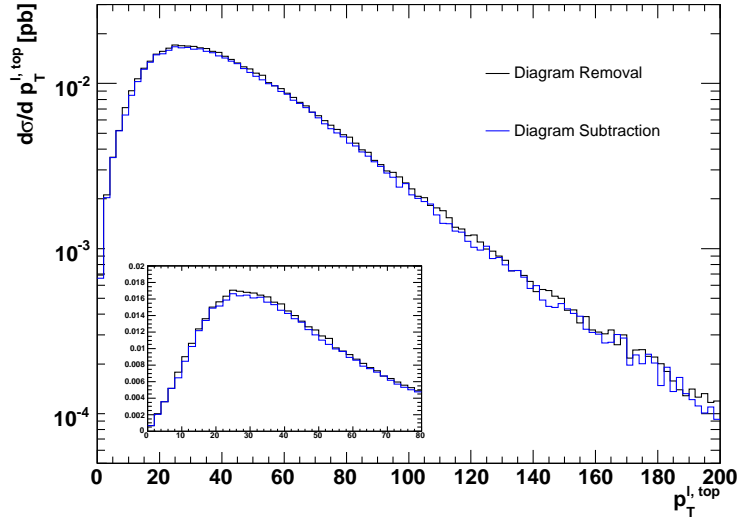,height=3.0in}
\caption{Comparison of DR and DS results, for the transverse momentum distribution of the lepton originating from the top decay.
\label{leptop}}
\end{center}
\end{figure}

In~\cite{Frixione:2008yi} a detailed comparison of DR and DS is made, where a transverse momentum veto on the second hardest $B$ jet is used as an example cut which can be used to reduce the $t\bar{t}$ background~\cite{Campbell:2005bb}. For phenomenologically reasonable values of this veto, the results from DR and DS agree closely, and certainly well within the scale variation associated with each result. As an example, the transverse momentum distribution of the lepton originating from the top decay (a worst-case amongst the distributions studied) is shown in figure~\ref{leptop}. This validates the statement that $Wt$ production can indeed be well-defined subject to cuts used to isolate the signal over the top pair background. When that is the case, MC@NLO can be used to provide an accurate description. One may generate samples of $Wt$ and $t\bar{t}$ events separately, and combine the two event samples consistently.

To summarise, we have implemented $Wt$ production in MC@NLO. This is non-trivial due to the interference between $Wt$ and top pair production. Thus two implementations of $Wt$ production have been given, whose difference measures the degree of interference with $t\bar{t}$. By running both codes, the user has a means of checking whether or not it is safe to consider $Wt$ to be well-defined. If this is indeed the case, one has a means of efficiently generating both $Wt$ and $t\bar{t}$ events, which can be consistently combined within existing systematic uncertainties. 

\section*{Acknowledgments}
This work was done in collaboration with Stefano Frixione, Eric Laenen, Patrick Motylinski and Bryan Webber. CDW was supported by the Dutch Foundation for Fundamental Research (FOM), and also by the Marie Curie Fellowship ``Top Physics at the LHC''.

\end{document}